\documentclass[12pt]{iopart}
\pdfoutput=1
\usepackage{iopams}
\usepackage{amsfonts}
\usepackage{bm}
\usepackage{graphicx}
\usepackage{epstopdf}
\usepackage[utf8x]{inputenc}
\usepackage{color}
\usepackage{subfigure}

\begin{document}

\title{Energy dissipation in an adaptive molecular circuit}

\author{Shou-Wen Wang$^{1,2}$, Yueheng Lan$^3$ and Lei-Han Tang$^{2,4}$}
\address{$^1$Department of Engineering Physics, Tsinghua University,
Beijing, 100086, China\\
$^2$Beijing Computational Science Research Center, Beijing, 100094, China\\
$^3$Department of Physics, Tsinghua University, Beijing, 100086, China\\
$^4$Department of Physics and Institute of Computational and Theoretical Studies, Hong Kong Baptist University, Hong Kong, China}

\begin{abstract}
 
The ability to monitor nutrient and other environmental conditions with high sensitivity is crucial for cell growth and survival. Sensory adaptation allows a cell to recover its sensitivity after a transient response to a shift in the strength of extracellular stimulus. The working principles of adaptation have been established previously based on rate equations which do not consider fluctuations in a thermal environment. Recently, G. Lan et al. (Nature Phys., 8:422-8, 2012) performed a detailed analysis of a stochastic model for the E. coli sensory network. They showed that accurate adaptation is possible only when the system operates in a nonequilibrium steady-state (NESS). They further proposed an energy-speed-accuracy (ESA) trade-off relation. 
We present here analytic results on the NESS of the model through a mapping to a one-dimensional birth-death process.  An exact expression for the entropy production rate is also derived.  Based on these results, we are able to discuss the ESA relation in a more general setting. Our study suggests that the adaptation error can be reduced exponentially  as
the methylation range increases.  Finally, we show that a nonequilibrium phase transition exists in the infinite methylation range limit, despite the fact that the model contains only two discrete variables.
\end{abstract}


\date{\today}

    
\section{Introduction}

As a paradigmatic example of environmental monitoring in biology,  the E. coli chemotactic sensory system has 
been studied extensively over the years~\cite{tu2013quantitative}.   Its core component is the 
transmembrane methyl-accepting chemotaxis protein (MCP) receptor. MCP binds selectively to 
ligands outside the cytoplasmic membrane and modulates the activity of its downstream 
signal transduction pathway in a way that depends on its methylation state. 
Two aspects are recognized to be crucial to the performance of the sensory network in the 
biological context: sensitivity of detection in a noisy environment, and adaptation to maintain 
that sensitivity over a broad range of ligand concentrations. With regard to high sensitivity to 
diffusing chemicals in the surrounding medium, Berg and Purcell~\cite{berg1977physics}   presented an optimal 
strategy in 1977 based on simple physical considerations. They showed that the measured 
chemotactic sensitivity of E. coli approaches that of the optimal design. Further indication of 
the organism’s optimal performance is found in its nearly perfect adaptation over five 
decades in ligand concentration. The latter property is shown to hold even when proteins on 
the sensory network are expressed away from their natural levels~\cite{alon1999robustness}. To explain this 
remarkable behavior, Barkai and Leibler (BL)~\cite{barkal1997robustness} introduced a simple model where 
the MCP methylation/demethylation rates are linked to the downstream activity. The system 
reaches a steady state only when its activity is at the level required by the balance of 
methylation and demethylation currents. It was soon pointed out by Yi et al.\cite{yi2000robust} that the BL 
scheme is in effect implementing an integral feedback control which is widely used in 
engineering systems to achieve robust adaptation. Furthermore, an exhaustive search by Ma et 
al. \cite{ma2009defining} to identify all possible 3-node adaptive networks found integral feedback control 
as one of only two core motifs that enable perfect adaptation.

In a separate development, there have been much progress in recent years in understanding fluctuation
phenomena in non-equilibrium systems whose dynamics do not satisfy detailed balance~\cite{jarzynski1997nonequilibrium,ritort2008nonequilibrium,sekimoto2010stochastic,harada2006energy, toyabe2010nonequilibrium,seifert2012stochastic}.  One particular aspect of fluctuations
in a nonequilibrium steady state (NESS) is the production of system's entropy and its subsequent
release as heat to the environment~\cite{esposito2007entropy}. In this respect, the generic behavior of 
nonequilibrium systems studied in the statistical physics community is shared by molecular processes
in a living cell. A well-known example is the kinetic proof-reading discussed by J. J. Hopfield in 1974~\cite{hopfield1974kinetic}. Here, the molecular machinery to carry out DNA replication can achieve a much lower error rate 
by operating out of equilibrium. It may be argued that employing energy flux to enable or improve the performance of a molecular circuit is a common practice in biology. However, there have been only a few examples so far
where the details are convincingly elucidated~\cite{qian2005nonequilibrium,mehta2012energetic, lan2013cost,sartori2014thermodynamic}.

The integral feedback control for adaptation requires asymmetric interactions between the 
output node and the integration node, which can only be realized by systems in a 
NESS. The issue of energy cost to maintain such a state was addressed 
in a recent study by Lan et al.~\cite{lan2012energy}. One of their main findings is a relation among the energy 
dissipation rate, adaptation speed and adaptation accuracy (ESA), which they 
suggested to hold generally. Their result is based on a sensory network model which has been shown to reproduce
most of the experimental data on the MCP receptor in E. coli~\cite{tu2013quantitative}.  

Despite its intuitive appeal, the ESA relation has not been derived from the more general results in the 
literature regarding the NESS.  Should there be a fundamental connection between the adaptation accuracy and 
energy dissipation rate?  Specifically,  is there a lower bound for energy dissipation rate to achieve a given 
adaptation accuracy?  To clarify this and other issues, it is necessary to perform a more comprehensive study 
of the sensory network model. Due to the conceptual importance of the sensory network model, 
a rigorous discussion is desirable.   

The paper is organized as follows. The biological background of adaptation and the model by Lan et al. are introduced in Section~\ref{sect:model}, followed by a detailed analysis of the NESS of the system in Section~\ref{sect:analysis}.  
In Section~\ref{sect:ESA} we drive an exact expression for the energy dissipation rate and compare it with the
ESA tradeoff relation. A nonequilibrium phase transition of the system in the infinite methylation range limit is identified in Section~\ref{sect:phaseTransition} and its properties discussed.  
Section~\ref{sect:conclusion} contains a summary of our results and conclusions. Mathematical details
of an approximate treatment of the NESS distribution is relegated to Appendix A.  

\section{A model for sensory adaptation}
\label{sect:model}

\begin{figure}[!ht]
\centering
\subfigure[]{\includegraphics[width=7cm]{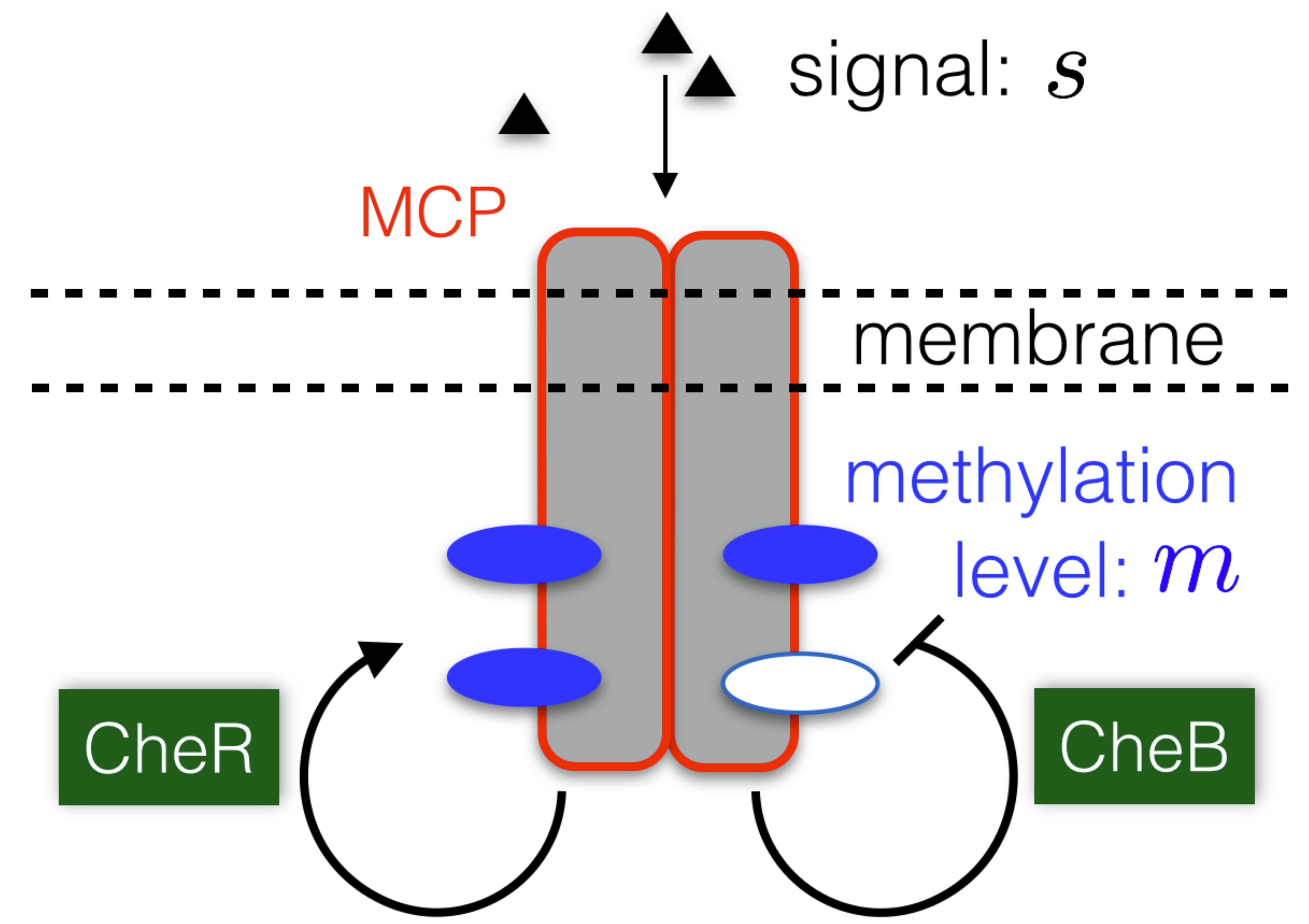}}
\subfigure[]{\includegraphics[width=7cm]{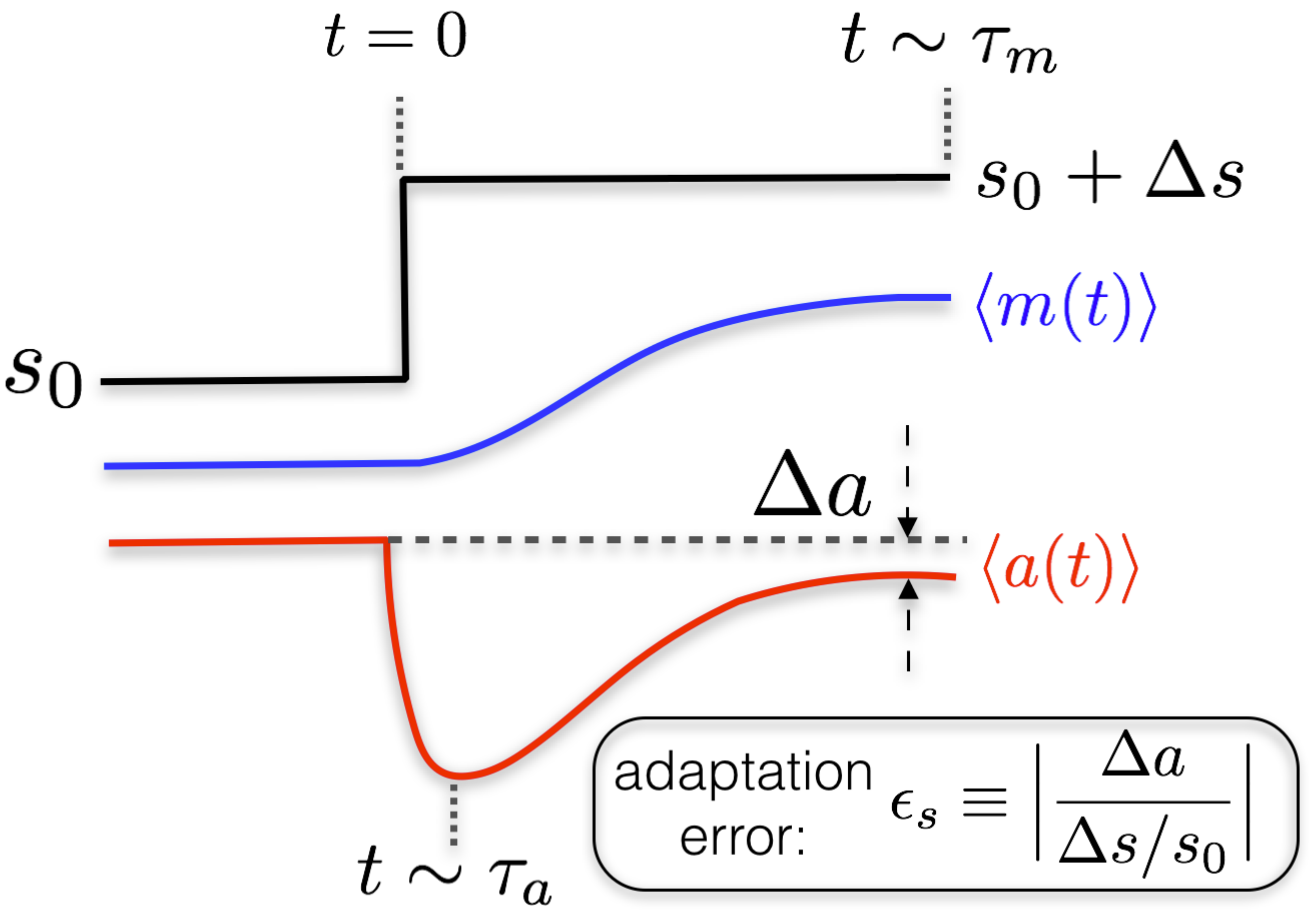}}
\caption{(a) The MCP receptor.   (b)  Mean response of the MCP receptor to a stepwise signal.  }
\label{fig:MCPmodel}
\end{figure}

Here, we briefly introduce the transmembrane methyl-accepting chemotaxis protein (MCP) receptor, which is the core component for receiving signal and exercising adaptation~\cite{tu2013quantitative}.   
MCPs regulate the clockwise-counterclockwise rotational switch of downstream flagellar motors which 
drive the run-and-tumble motion of an E. coli cell.   
A simple cartoon of this receptor is illustrated in Figure~\ref{fig:MCPmodel}(a).   
The activity of the receptor can be described by a binary variable $a$: $a=1$ for the active state and $a=0$ 
for the inactive state. The transition rate between the two states 
depends on the external ligand concentration (i.e., signal strength) $s$ and the internal methylation level $m$.   
$m$ ranges from $0$ to $m_0$,  with $m_0=4$ for a single MCP.   
The methylation level can be increased by enzyme CheR and decreased by enzyme CheB,  
in a way that depends on the activity of MCP.    
Figure~\ref{fig:MCPmodel}(b) illustrates response of the MCP receptor to a stepwise signal $s$ obtained from
experimental measurements.   
The mean activity $\langle a(t)\rangle$ changes sharply in a short time window $\tau_a$ less than a second, and recovers slowly over a much longer time scale $\tau_m$, of the order of a minute, due to the slow change of average internal methylation level $\langle m(t)\rangle$. The output recovery after a transient response to external stimuli is called adaptation.  The performance of adaptation is characterized by the adaptation error which can be defined as the ratio between
the final shift of activity and the relative change of signal strength, as illustrated in Figure~\ref{fig:MCPmodel}(b).  

\begin{figure}[!ht]
\centering
\includegraphics[width=14cm]{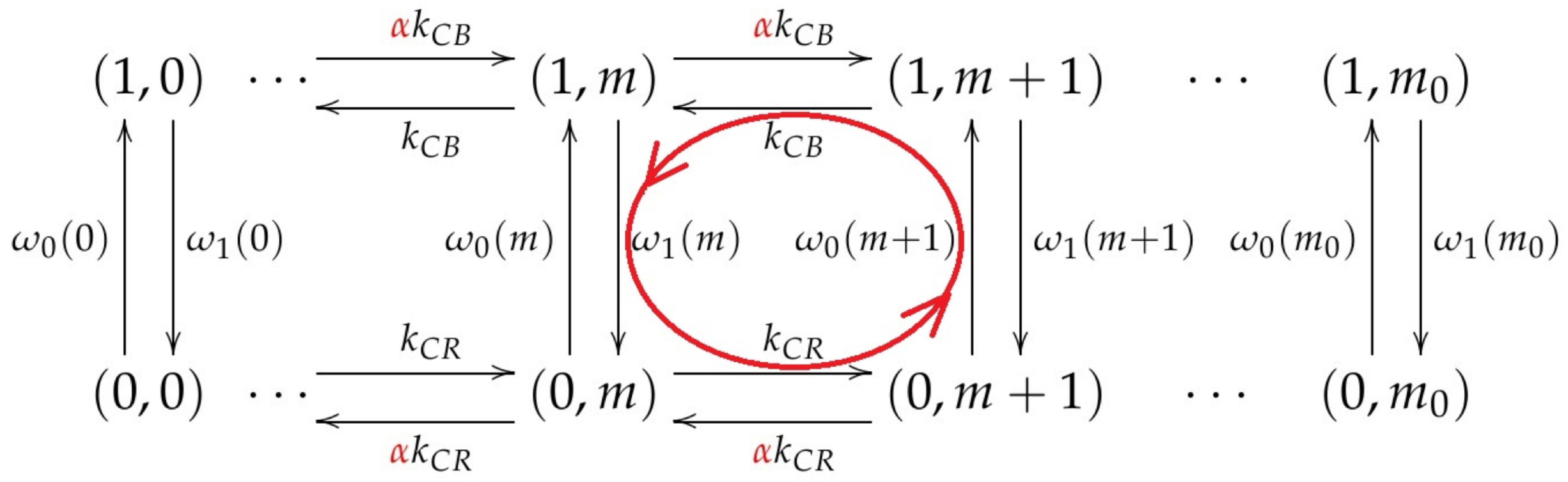}
\caption{The Markov network model of a single receptor MCP in E. coli.  
Red arrows indicate existence of a {\it futile cycle} at small $\alpha$, 
which is essential for adaptation. }
\label{fig:model}
\end{figure}

A Markov network model with internal states specified by $(a,m)$ was proposed for a single MCP by 
Lan et al.~\cite{lan2012energy},  as shown in Figure~\ref{fig:model}.   
Transition between active and inactive states at a given methylation level takes place on the time scale $\tau_a$,  
with rates given by 
\begin{equation*}
\omega_1(m,s)=\frac{1}{\tau_a} \exp\Bigl(\frac{\beta}{2}\Delta E(m,s)\Bigr),\qquad  
\omega_0(m,s)=\frac{1}{\tau_a} \exp\Bigl(-\frac{\beta}{2}\Delta E(m,s)\Bigr), 
\end{equation*}
where $\beta=1/T$ is the inverse temperature, and 
\begin{equation*}
\Delta E(m,s)=e_0 (m_1-m)+f(s)
\end{equation*} 
the free energy difference between states $(0,m)$ and $(1,m)$, with  $f(s)=\ln [(1+s/K_i)/(1+s/K_a)]$.
Here $e_0>0$ is the methylation energy, $m_1$ an offset methylation level~\cite{lazova2011response,tu2013quantitative},
and $K_i$ and $K_a (\gg K_i)$ are equilibrium constants for ligand binding to the receptor in the inactive and active states,
respectively.  
Transition between different methylation levels takes place on the time scale $\tau_m$, with rates 
indicated in Figure~\ref{fig:model}: when the receptor is inactive, the rate of methylation (assisted by enzyme 
CheR) is $K_{CR}$ while the rate of demethylation is $\alpha K_{CR}$; when the receptor is active, the rate of
demethylation (assisted by CheB) is $K_{CB}$ while the rate of methylation is $\alpha K_{CB}$.
An estimate of the typical methylation/demethylation cycle time is given by $\tau_m=k_{CB}^{-1}+k_{CR}^{-1}$.    

The parameter $\alpha$ specifies the degree of disequilibrium in the system. When 
$\alpha=\alpha_{EQ}\equiv\exp(\beta e_0/2)$, global detailed balance condition is satisfied, in which case
the Boltzmann distribution governed by a free energy function is recovered.     
For $\alpha<\alpha_{EQ}$, the system is driven out of equilibrium with generally different properties which we
study using both analytical and numerical methods. Therefore $\alpha$ describes the strength of driving
to keep the receptor to operate under out of equilibrium conditions.

In the numerical examples presented below, we adopt the parameter values as
suggested in Ref. \cite{lan2012energy}: $m_1=1$,  
$K_i=18.2 \mu$M, $K_a=3000 \mu$M, $\beta=1$ ($kT$ as the unit of energy) and $e_0=2$. 
The time constants are chosen as $\tau_a=0.1 s$, and $k_{CB}=k_{CR}=0.01 s^{-1}$.   
For this parameter set,  $\alpha_{EQ}=e$. 
To simplify the notation, we write $\omega_1(m,s), \omega_0(m,s)$ as  $\omega_1(m)$ and $\omega_0(m)$ respectively.  

\section{Adaptation and the NESS}
\label{sect:analysis}

\subsection{Condition for adaptation}

We first revisit the condition for adaptation first obtained by Lan et al.~\cite{lan2012energy}. 
For the model introduced in Sec. 2, the master equation for the joint probability $P(a,m)$ takes the form,
\numparts
\begin{eqnarray}
\frac{dP(0,m)}{dt}&=&k_{CR}P(0,m-1) +\alpha k_{CR} P(0,m+1) +
 \omega_1(m) P(1,m)\nonumber\\
 && - [k_{CR}+\alpha k_{CR}+\omega_0(m)] P(0,m),
\label{eq:govern1}\\
\frac{dP(1,m)}{dt}&=&\alpha k_{CB} P(1,m-1) +k_{CB} P(1,m+1) +\omega_0(m) P(0,m) \nonumber\\
&&- [\alpha k_{CB}+k_{CB}+\omega_1(m)]P(1,m).
\label{eq:govern2}
\end{eqnarray}
\endnumparts
From the above, we obtain the evolution equations for the moments
$\langle m\rangle=\sum_{a,m} mP(a,m)$ and $\langle a\rangle=\sum_m P(1,m)$:
\numparts
\begin{eqnarray}
\frac{d\langle a\rangle }{d t} &=&\sum_m \Big[ \omega_0(m)P(0,m)-\omega_1(m)P(1,m)\Big],\\
\frac{d\langle m\rangle}{dt}&=&(1-\alpha) (k_{CR}+ k_{CB})\Bigl(-\langle a\rangle +\frac{k_{CR}}{k_{CR}+k_{CB}}\Bigr)+B_1.
  \label{eq:mrate}
  \end{eqnarray}
\endnumparts
Here $B_1= \alpha k_{CR} P(0,0) +k_{CB} P(1,0) - k_{CR} P(0,m_0) - \alpha k_{CB} P(1,m_0)$
depends on the probabilities for the extreme methylation states $m=0$ and $m=m_0$.   

\begin{figure}[!ht]
\centering
 \includegraphics[width=7cm]{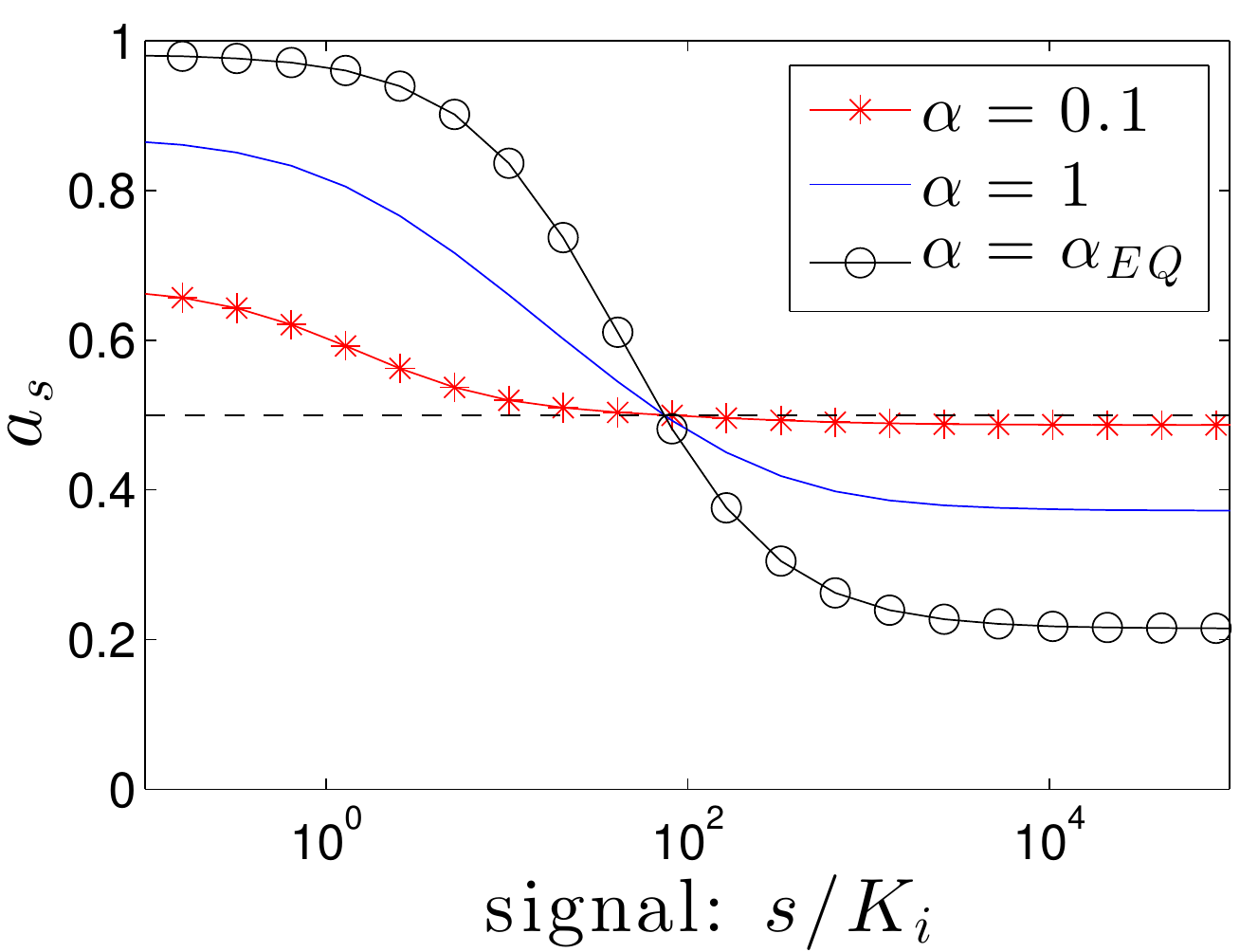}
 \caption{Mean receptor activity against ligand concentration for three different values of the nonequilibrium 
 parameter $\alpha$.  The dash line indicates the value $a_0$.  Here the methylation range $m_0=4$.}
 \label{fig:signalDependence}
 \end{figure}

In a steady environment of constant ligand concentration $s$, the system is expected to reach a steady state in a time
$\tau_m$ where both $\langle m\rangle$ and $\langle a\rangle$ assume constant values.
Setting the right-hand-side of Eq. (\ref{eq:mrate}) to zero yields,
\begin{equation}
\langle a\rangle=a_s=\frac{k_{CR}}{k_{CR}+k_{CB}}+\frac{1}{1-\alpha}\frac{B_1}{k_{CR}+k_{CB}}.
\label{eq:avera}
\end{equation}
Since the methylation and demethylation rates $k_{CR}$ and $k_{CB}$ are
assumed to be constants in the model, the first term $a_0\equiv k_{CR}/(k_{CR}+k_{CB})$ on
the right-hand-side of Eq. (\ref{eq:avera}) is independent of $s$.
Figure~\ref{fig:signalDependence} shows $a_s$ against $s$ for three different values of $\alpha$,
obtained from numerically exact solution of the model in the NESS.   
The steady-state activity $a_s$ is centered around $a_0$ (dashed line) over a large range of $s$ for $\alpha<1$, 
but not so for $\alpha\ge 1$.  The ``adaptation error''
\begin{equation}
\epsilon\equiv  |a_s -a_0  |=\Big|\frac{1}{1-\alpha}\frac{B_1}{k_{CR}+k_{CB}}\Big|
\label{eq:error}
\end{equation}
is essentially controlled by the size of the boundary term $B_1$.
For $\alpha<1$, $B_1$ is small over a broad range of $s$. As we shall see in the next section, the NESS distribution
in this case is indeed centered in the middle of the allowed methylation range. This is however not the case when $\alpha>1$. 

\begin{figure}[!ht]
\centering
\subfigure[]{\includegraphics[width=7cm]{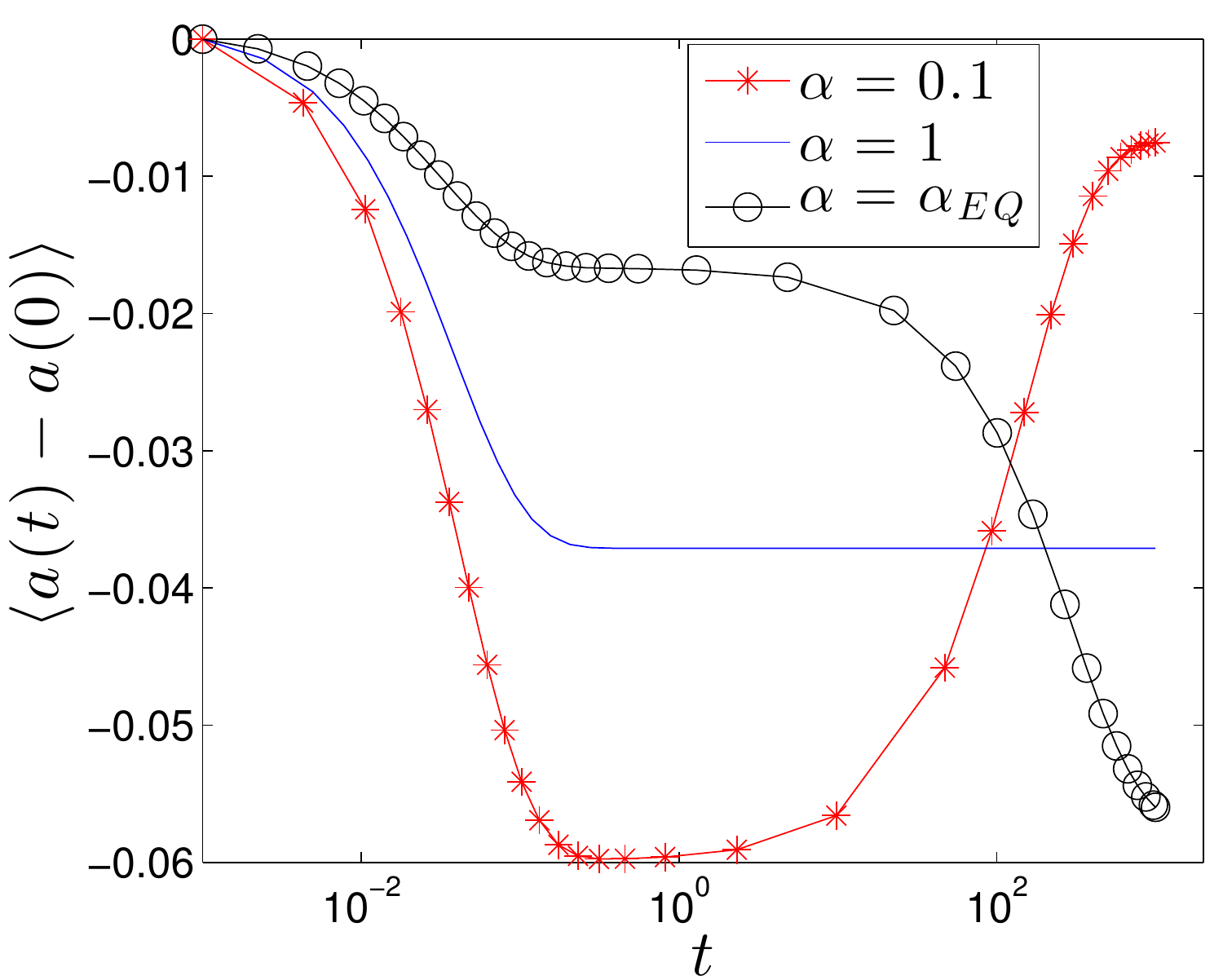}}
\subfigure[]{\includegraphics[width=7cm]{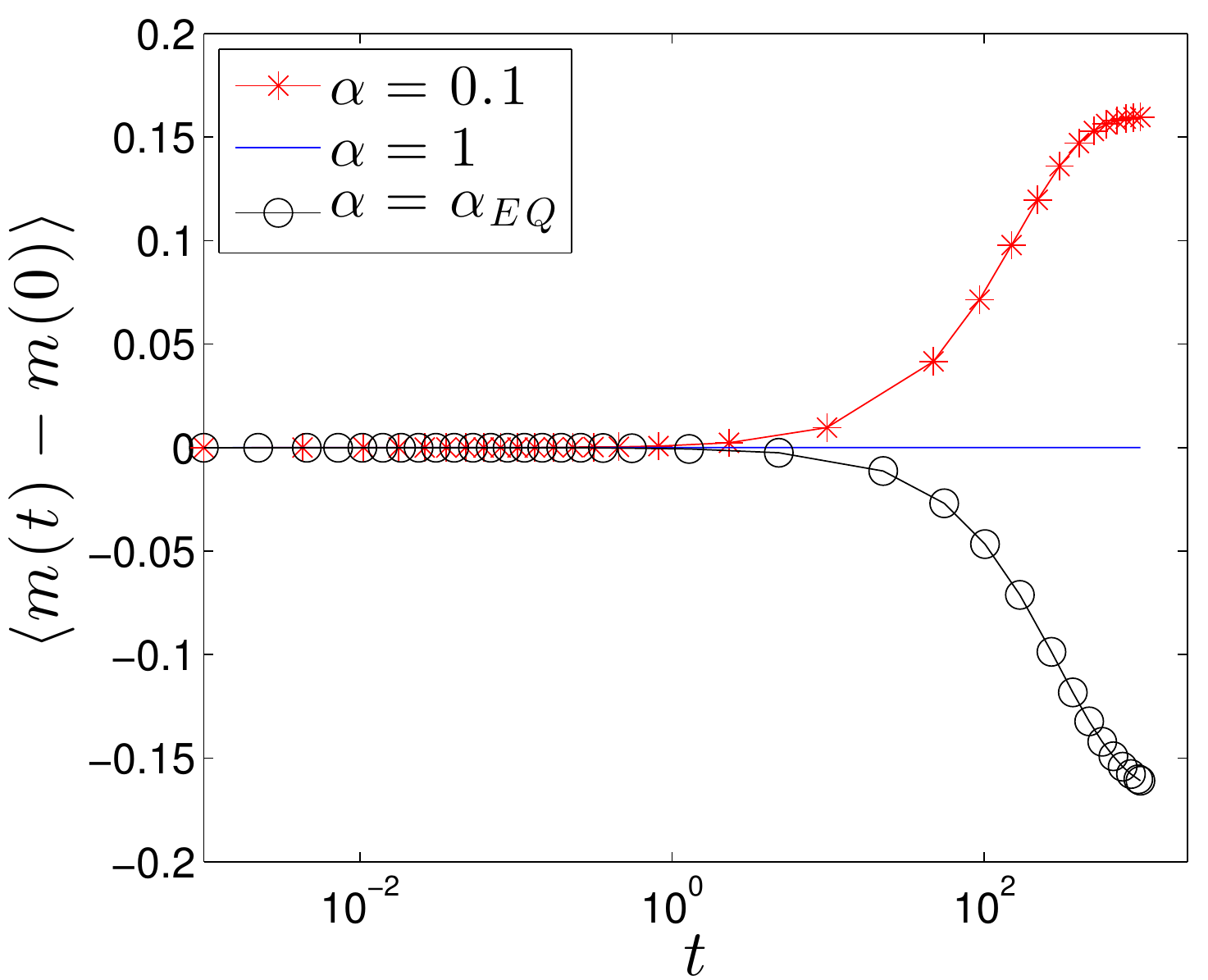}}
\caption{An initial steady state at $s=10K_i$ is perturbed by shifting the ligand concentration to $s=15K_i$ at $t=0$.  
Results at three different values of $\alpha$ are shown:
(a) change in the mean receptor activity $\langle a(t)\rangle$ against $t$; (b) change in the methylation level $\langle m(t)\rangle$ 
against $t$.  The methylation range $m_0$ is set to be four.}
\label{fig:signalChangeResponse}
\end{figure}

The transient response to a signal ramp also exhibits qualitatively different behavior for $\alpha<1$ and $\alpha>1$. 
Figure~\ref{fig:signalChangeResponse} shows our results obtained by numerically integrating the master
equations (\ref{eq:govern1}) and (\ref{eq:govern2}) at three different values of $\alpha$,
upon a jump in ligand concentration from $10K_i$ to $15K_i$ at $t=0$. 
The initial response of $\langle a\rangle$ to the signal ramp is qualitatively similar in the three cases, 
i.e., a fast depression of receptor activity to a near plateau value in a time of order $\tau_a$. 
However, opposite behavior is seen at longer times, in
concert with the change in methylation level as seen in Fig. ~\ref{fig:signalChangeResponse}(b).
For $\alpha>1$, a further decrease of the mean activity is seen when the methylation level starts to decrease in
response to the change in $\langle a\rangle$. On the other hand, when $\alpha<1$, the methylation level
increases in accordance with Eq. (\ref{eq:mrate}), eventually restoring 
the mean activity to a value close to the pre-stimulus level. The latter is precisely the scenario for 
adaptation that employs a change in the methylation level to offset the activity change effected by
the shift in signal strength.  

In summary, both the steady-state activity and the transient response to a signal ramp show qualitatively different
behavior below and above $\alpha=1$. 
We thus conclude that the condition for adaptation in this model is $\alpha<1$.

\subsection{The NESS distribution}

In the previous subsection, we obtained the condition for adaptation by considering the moment equations with the
help of numerical integration of the master equation. To gain a complete understanding of the NESS, it is necessary
to calculate the distribution function $P(a,m)$. Fortunately, for the model in question, this can be done under the
``fast equilibrium'' approximation facilitated by the separation of the time scales $\tau_a$ and $\tau_m\gg\tau_a$.
Then,  $P(1,m)$ and $P(0,m)$ satisfy the local detailed balance 
\begin{equation}
\frac{P(1,m)}{P(0,m)}= \frac{\omega_0(m)}{\omega_1(m)}+O(\tau_a/\tau_m),\qquad (0\leq m\leq m_0).
\label{eq:localDetailedBalance}
\end{equation}
Let $P(m)\equiv P(0,m)+P(1,m)$, we obtain,
\begin{equation}
\fl 
P(0,m)=\frac{1}{1+\exp[-\beta\Delta E(m,s)]}P(m),\qquad 
P(1,m)=\frac{\exp[-\beta\Delta E(m,s)]}{1+\exp[-\beta\Delta E(m,s)]} P(m).
\label{eq:fast-eq}
\end{equation}
With the help of (\ref{eq:localDetailedBalance}), Eqs. (\ref{eq:govern1}) and (\ref{eq:govern2}) combine to yield  
\begin{equation}
\frac{dP(m)}{dt}= b(m-1) P(m-1)+d(m+1) P(m+1)-[b(m)+d(m)] P(m).
\label{eq:masterSeparation}
\end{equation}

Equation (\ref{eq:masterSeparation}) defines a one-dimensional birth-death process 
with the birth and death rates given respectively by,
\[\fl b(m)= \frac{k_{CR}+\alpha k_{CB}  \exp[-\beta \Delta E(s,m)] }{ 1+\exp[-\beta  \Delta E(s,m)]},\qquad 
d(m)=\frac{ \alpha k_{CR} + k_{CB} \exp[-\beta\Delta E(s,m)] }{1+\exp[-\beta\Delta E(s,m)]}.\]
Its steady-state distribution takes the form,
\begin{equation}
P(m+1)=\frac{b(m)}{d(m+1)}P(m)=P(0)\prod_{i=0}^{m} \frac{b(i)}{d(i+1)}.
\label{eq:coar-PDF}
\end{equation}
Together with Eq. (\ref{eq:fast-eq}) the full NESS distribution is obtained.

\begin{figure}[!ht]
\centering
\subfigure[]{\includegraphics[width=7cm]{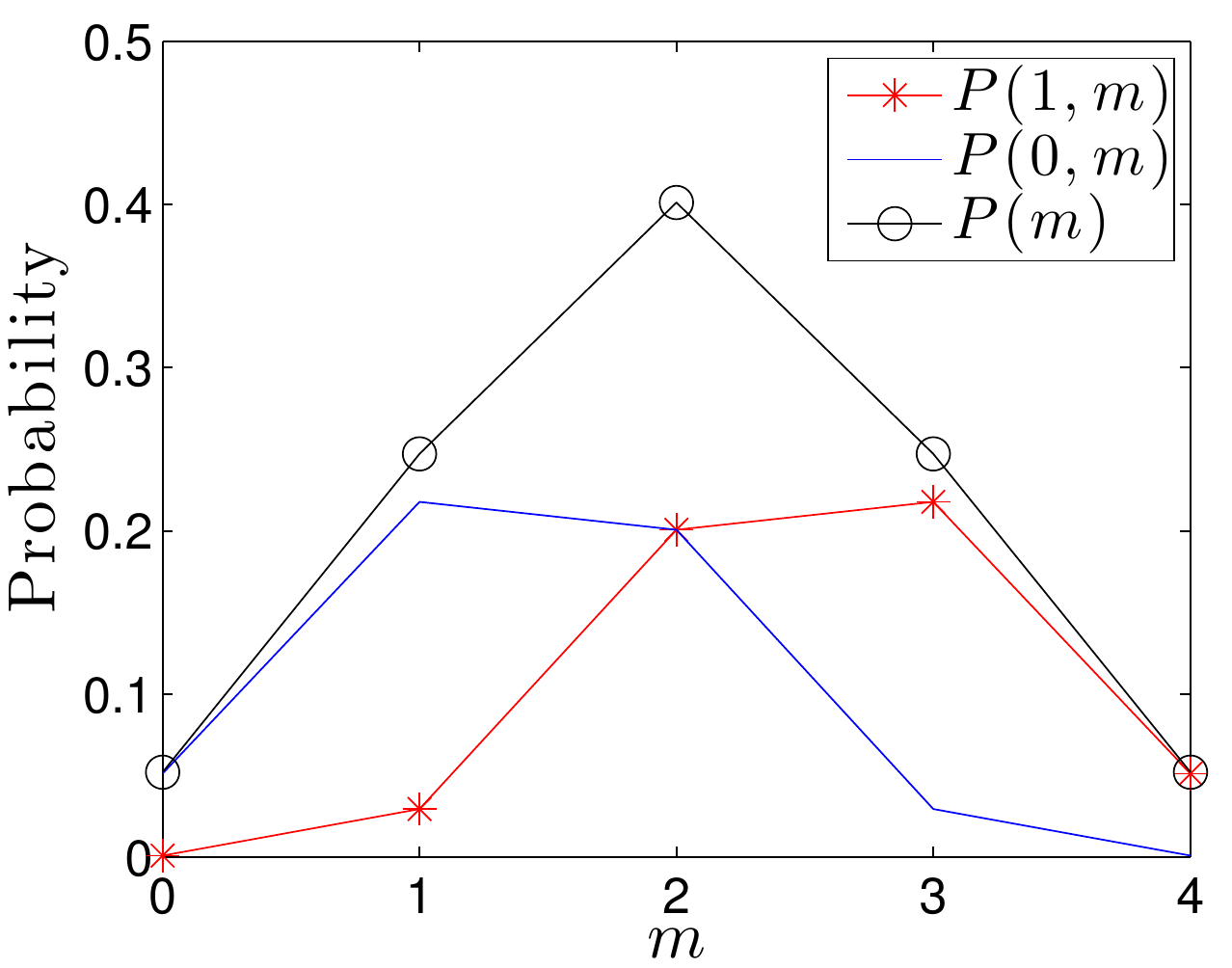}}
\subfigure[]{\includegraphics[width=7cm]{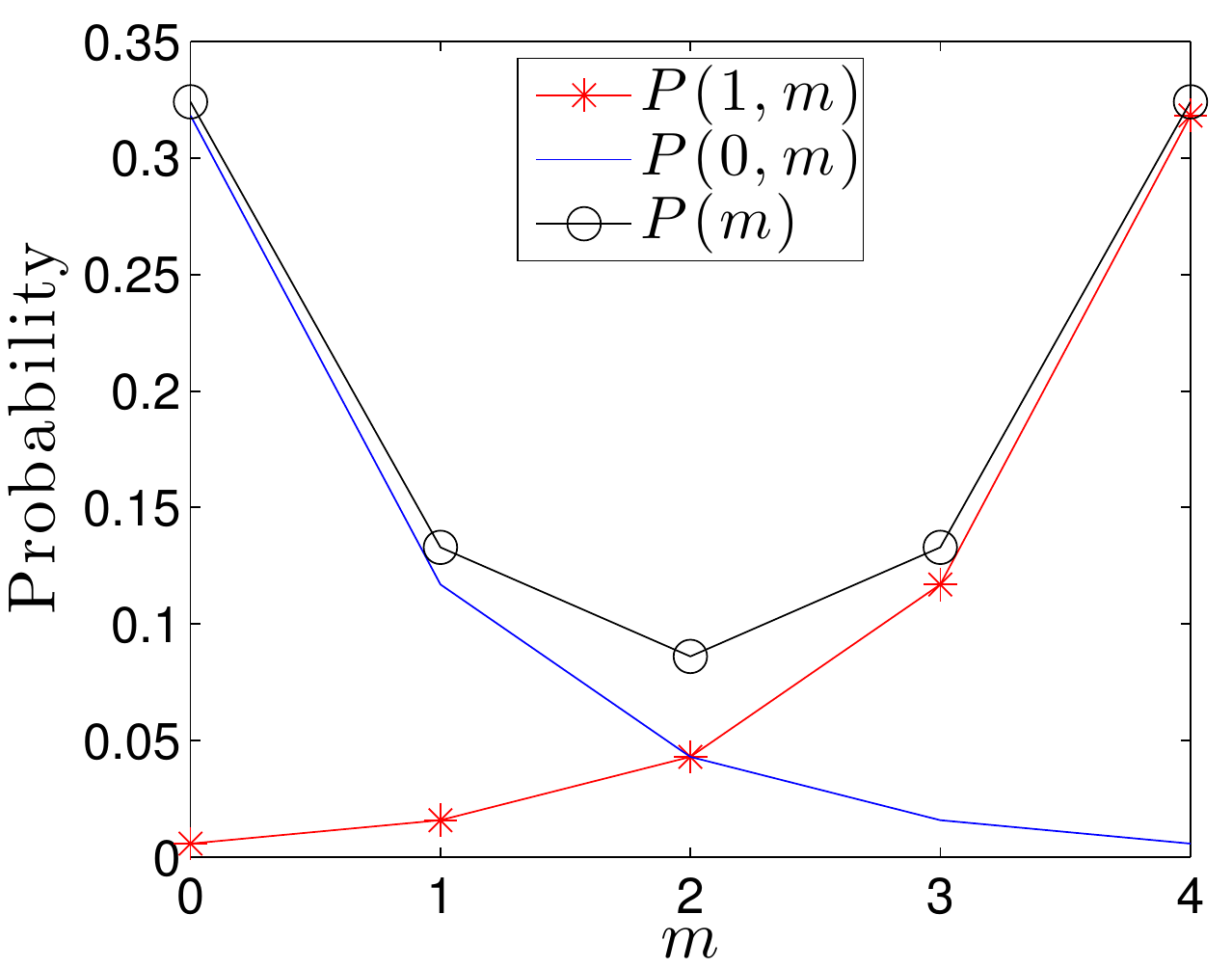}}
\caption{The steady-state distributions $P(1,m)$,  $P(0,m)$ and $P(m)$ at $m_0=4$,  $s=10 K_i$.
(a) $\alpha=0.1$; (b) $\alpha=\alpha_{EQ}>1$.   }
\label{fig:distribution}
\end{figure}

Consider the range of ligand concentrations where the receptor is functional, i.e.,
$\exp[-\Delta E(m,s)]\ll 1$ at $m=0$ (inactive state favored) and $\exp[-\Delta E(m,s)]\gg 1$ at $m=m_0$ 
(active state favored).  Consequently, the ratio $b(m)/d(m+1)$ changes monotonically between the limiting
values $1/\alpha$ and $\alpha$ as $m$ increases from $0$ to $m_0$. Let $m^*$ be the value of $m$ where
$b(m^*)/d(m^*+1)\simeq 1$. According to Eq. (\ref{eq:coar-PDF}), this is the methylation level where 
$P(m)$ varies slowest with $m$, i.e., the stationary point of the distribution.
For $\alpha<1$, $b(m)/d(m+1)>1$ on the low methylation side ($m<m^*$) while $b(m)/d(m+1)<1$ on the
high methylation side ($m>m^*$). Therefore $P(m)$ reaches its peak value at $m^*$.
The opposite situation happens for $\alpha>1$, where $P(m)$ initially decreases with $m$ on the low methylation
side, reaches its minimum value at $m^*$, and increases on the high methylation side.   
At $\alpha=1$, $b(m)=d(m)$ so that $P(m)=P(0)b(0)/d(m)$ becomes essentially flat especially when $k_{CR}=k_{CB}$.
The general behavior of the NESS distributions in the two regimes are illustrated in Figure~\ref{fig:distribution}.  
A more complete discussion of the functional form of these distributions and their dependence on $s$ at different $\alpha$
values can be found in the Appendix A.

In the case $\alpha<1$, the signal level affects the shape of the distribution by shifting its peak position 
$m^*= m_1+f(s)/e_0-(\beta e_0)^{-1}\ln(k_{CB}/k_{CR})$ (see Appendix A) which coincides with the mean methylation level.   
This is a general feature for adaptation achieved through integral feedback control, i.e., 
the effect of external signal change is absorbed by a shift in the average methylation level.

\subsection{Adaptation error}

According to  Eq. (\ref{eq:avera}), the mean receptor activity $\langle a\rangle$ in the NESS depends on the
signal level $s$ only through the probabilities for the extreme methylation states at $m=0$ and $m=m_0$.  
For $\alpha<1$, $P(m)$ decreases rapidly away from the peak position at $m^\ast$.
Except very close to $\alpha = 1$ which requires a separate treatment, the adaptation error as defined by 
Eq. (\ref{eq:error}) can be estimated by the largest term in the expression for $B_1$.
Let $m_d\equiv {\rm min}\{m^*, m_0-m^*\}$ be the distance between $m^\ast$ and the closest extreme methylation state.
With the help of Eq. (\ref{eq:probability-active}), we obtain, 
\begin{equation}
\ln\epsilon\simeq
\left\{
\begin{array}{lc}
-{1-\alpha\over 1+\alpha}{\beta e_0\over 2}m_d^2, & m_d < {\alpha +1\over\alpha -1}{1\over \beta e_0}\ln\alpha;\\
m_d\ln\alpha +{1+\alpha\over 1-\alpha}{1\over 2\beta e_0}\ln^2\alpha, & {\rm otherwise.}
\end{array}\right. 
\label{eq:errorApprox}
\end{equation}

Equation (\ref{eq:errorApprox}) shows that the adaptation error can be decreased by
either increasing the methylation range $m_0$ or decreasing the parameter $\alpha$ that brings
the system further away from equilibrium. At a given $\alpha<1$, increasing
$m_0$ allows a greater functional range of the receptor and consequently larger values for $m_d$,
resulting in an exponential decrease of $\epsilon$.
On the other hand, at a fixed $m_0$, decreasing $\alpha$ increases the rate of exponential decay of $\epsilon$.
However, when $\alpha$ is below $\alpha_m=\exp(-\beta e_0 m_0/2)$, the error basically saturates
to a value bounded from below by $\epsilon_m\simeq \exp(-\beta e_0 m_0^2/8)$.
These observations are in agreement with the trends see in 
Fig.~\ref{fig:ESArelation}(a) where $\epsilon$ is plotted against $\alpha$ for several different values of $m_d$. 

\section{Energy dissipation and the ESA trade-off}
\label{sect:ESA}

The nonequilibrium methylation/demethylation dynamics of the MCP receptor requires energy 
input~\cite{lan2012energy}. Within the adaptation model considered here, the rate of energy dissipation can
be calculated using the standard formula~\cite{lebowitz1999gallavotti, seifert2005entropy, qian2007phosphorylation}
\begin{equation}
 \dot{W}=\frac{1}{2\beta} \sum_{X,X'} J(X|X')\ln \frac{\omega(X|X')}{\omega(X'|X)}. 
 \label{eq:W-dot}
 \end{equation}
Here  $\omega(X|X')$ is the transition rate from state $X'$ to state $X$,  $J(X|X')=\omega(X|X') P(X')-\omega(X'|X)P(X)$ 
is the net flux from $X'$ to $X$, and $P(X)$ is the probability for state $X$.
For our purpose, it is convenient to rewrite the above equation in terms contributions from 
directed ``elementary cycles'' ~\cite{andrieux2007fluctuation}.  
An elementary cycle is a loop formed by nodes and edges of the network
that cannot be further decomposed into smaller loops.
Denoting by $C_l$ the $l$th elementary cycle on the network,  Eq. (\ref{eq:W-dot}) can be rewritten as
\begin{equation}
\dot{W}=\frac{1}{\beta} \sum_l  J(C_l)\mathcal{A}(C_l),
\label{eq:W-dot-cycle}
\end{equation}
where $J(C_l)$ is the probability flux associated with cycle $C_l$,  
and $\mathcal{A}(C_l)=\sum_{e\in C_l} \ln [\omega(X'|X)/\omega(X|X')]$, 
summed along the cycle.    

For the network model shown in Fig.~\ref{fig:model}, we define the $m$th elementary cycle
to be the rectangle between methylation levels $m$ and $m+1$, directed counter-clockwise as
indicated by the red arrows.
It is simple to verify that the thermodynamic force $\mathcal{A}(C_l)=2\ln [\alpha_{EQ}/\alpha]$ is the same for all cycles.
The cycle flux $J(m)=k_{CB}P(1,m+1)-\alpha k_{CB}P(1,m)$ can also be read off easily from the figure.
From Eq. (\ref{eq:W-dot-cycle}) we then obtain,
\begin{eqnarray*} 
\dot{W}&=&{2\over\beta}\Bigl(\ln \frac{\alpha_{EQ}}{\alpha}\Bigr)\sum_{m=0}^{m_0-1}[k_{CB}P(1,m+1)-\alpha k_{CB}P(1,m)]\\
&=&{2k_{CB}\over\beta}\Bigl(\ln \frac{\alpha_{EQ}}{\alpha}\Bigr)
[(1-\alpha)a_s-P(1,0)+\alpha P(1,m_0)].
\end{eqnarray*}
With the help of Eq. (\ref{eq:avera}),  we obtain finally the following exact expression for the energy dissipation
in the NESS,
\begin{equation}
\dot{W}=\frac{2}{\beta \tau_m} \Big( 1-\alpha +B_2 \Big)\ln \frac{\alpha_{EQ}}{\alpha} , 
\label{eq:exactW}
\end{equation}
where $B_2=\alpha P(0,0)-P(1,0) +\alpha P(1,m_0) -P(0,m_0)$ is a boundary term.
    
\begin{figure}[!ht]
\centering
\subfigure[]{ \includegraphics[width=7cm]{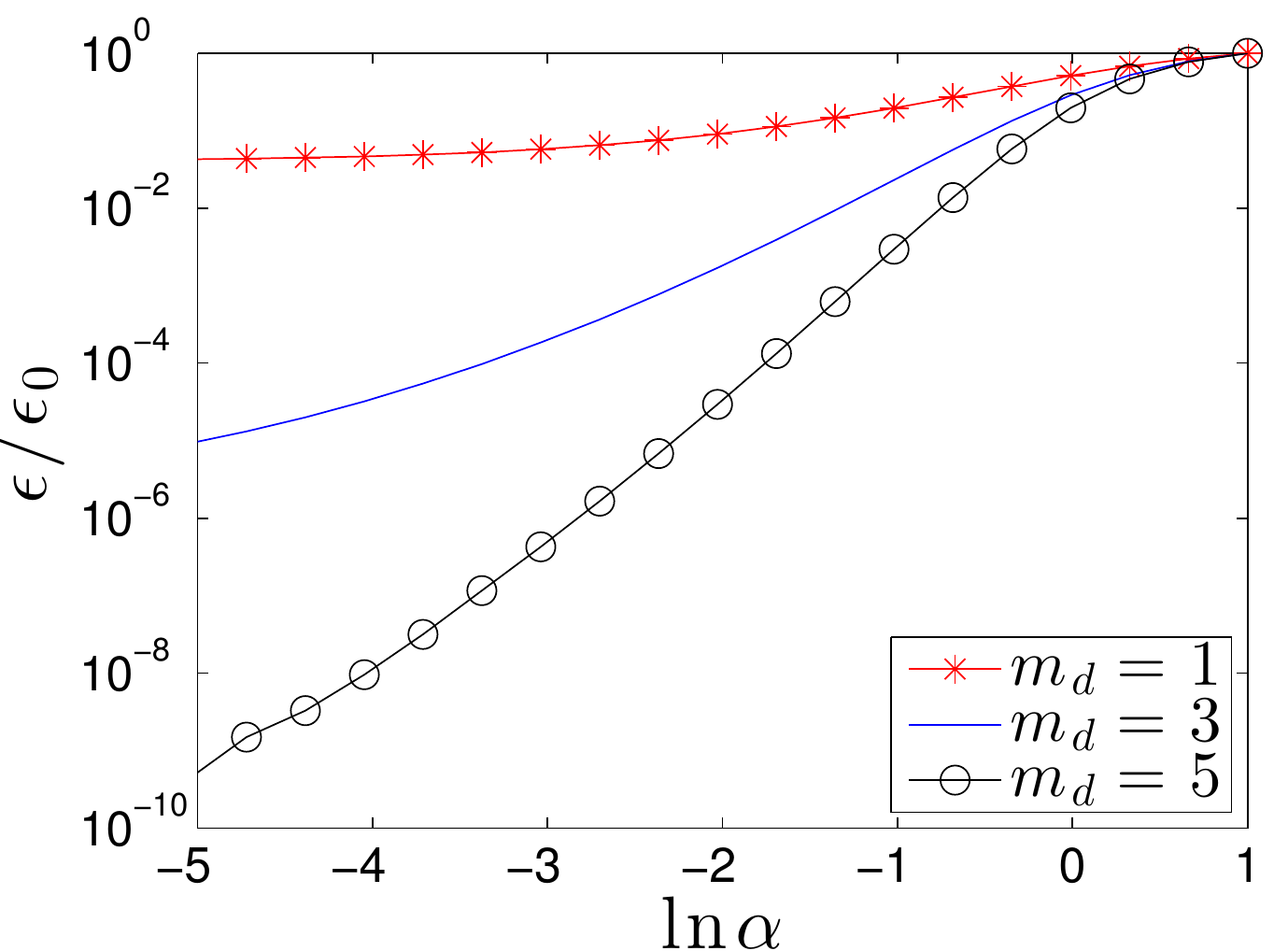}}
\subfigure[]{\includegraphics[width=7cm]{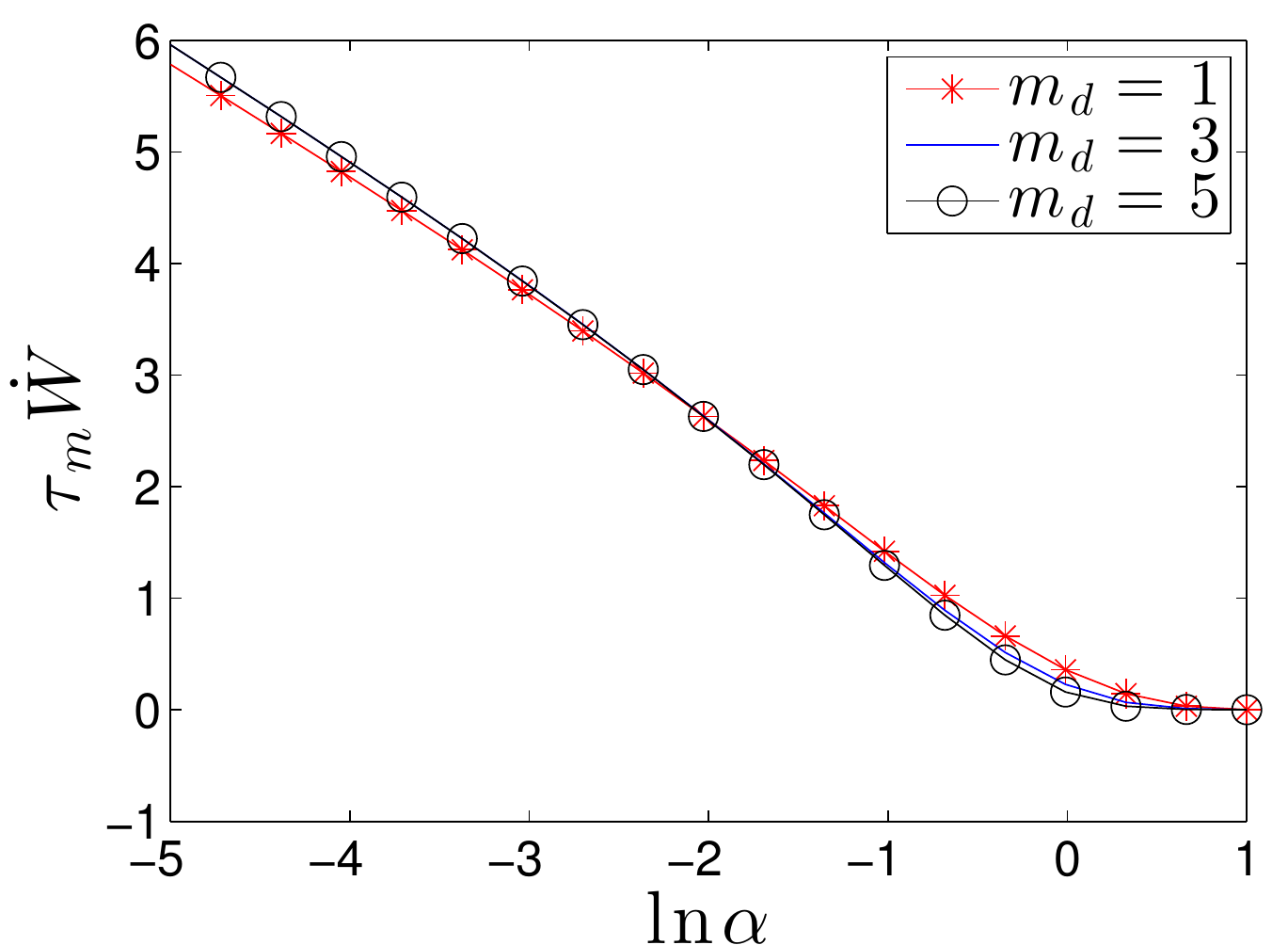}}
\caption{ (a)  Adaptation error and (b) energy dissipation over the methylation time scale $\tau_m$ against
$\ln\alpha$.  Here the ligand concentration is chosen such that $m_d=m^*=m_0/2-1$ with
$m_0=4,\;8$ and $12$.}
\label{fig:ESArelation}
\end{figure}

Figure~\ref{fig:ESArelation}(b)  shows $\dot{W}$ against $\alpha$ for selected values of $m_d$.
In all cases presented, a logarithmic increase on the far-from-equilibrium side (i.e., $\alpha\ll 1$)
is seen, in agreement with Eq. (\ref{eq:exactW}). Dependence of $\dot{W}$ on $m_d$, which enters only through
the boundary term $B_2$, is essentially negligible. This behavior can be understood from the fact that
most of the dissipation takes place in the loop centered around the peak position $m^\ast$ of the NESS
distribution $P(m)$.

In Ref.~\cite{lan2012energy}, based on approximate solutions of the adaptation model at $m_0=4$,
Lan et al. proposed the Energy-Speed-Accuracy tradeoff relation,
\begin{equation}
\dot{W}\simeq (c_0\sigma_a^2)\omega_m\ln(\epsilon_0/\epsilon)
\label{eq:ESA}
\end{equation}
to capture the increase in energy dissipation to achieve higher accuracy of adaptation as $\alpha$ is reduced.
Here $c_0\sigma_a^2$ sets the appropriate energy scale for the problem, and $\omega_m=\tau_m^{-1}$.
Comparing with our results Eqs. (\ref{eq:errorApprox}) and (\ref{eq:exactW}), 
we see that Eq. (\ref{eq:ESA}) needs to be modified to take into account the dependence of $\epsilon$
on the distance $m_d$ from the actual mean methylation level to the boundaries of the full methylation
range, i.e., $m=0$ and $m=m_0$. In addition, as we see from Fig.~\ref{fig:ESArelation} , the adaptation
error saturates to a value of the order of $\epsilon_m$ set by $m_0$ when $\alpha$ falls below $\alpha_m$, while
the energy dissipation rate $\dot{W}$ keeps increasing. From the calculations presented above, we see that 
$\epsilon$ is controlled by the probabilities for the rare events where the extreme methylation states 
are visited, while $\dot{W}$ is not sensitive to the actual methylation level itself.

\section{Phase transition}
\label{sect:phaseTransition}

As we have seen in Sec. 3, there is a qualitative change in the shape of the NESS distribution $P(m)$ at $\alpha=1$.
For $\alpha>1$, $P(m)$ is bimodal with peaks at the two ends of the methylation range from $0$ to $m_0$. 
The relative weight of the peaks is controlled by the signal strength $s$.
On the other hand, for $\alpha<1$, $P(m)$ has a single peak in the middle of the methylation range. As the signal
strength $s$ varies, the peak position shifts accordingly but its shape remains more or less the same until either
end of the methylation range is reached. As discussed previously by Lan et al.~\cite{lan2012energy}, the latter feature 
is crucial for the implementation of precise adaptation. In this section we examine the transition
between the two regimes in further detail with the help of the exact solution for the NESS.

Let us first examine the behavior of the energy dissipation rate in the NESS given by Eq. (\ref{eq:exactW})
in the limit $m_d=\min\{m^\ast,m_0-m^\ast\}\rightarrow\infty$.
For $\alpha<1$, Eq. (\ref{eq:probability-active}) shows that the boundary probabilities vanish in this limit,
and hence $B_2=0$.  On the other hand, for $\alpha>1$, 
this limit implies the activation energies $\Delta E(0,s)\rightarrow +\infty$ and $\Delta E(m_0,s)\rightarrow -\infty$.
Therefore the receptor is nearly exclusively in the inactive state when the methylation level is close to zero,
and exclusively in the active state when the methylation level is close to full. Then, the elementary loop
current $J(m)=0$ for all $m$ which in turn yields vanishing dissipation. Here we encounter an interesting example
where the detailed balance is violated by the kinetic rates but no dissipation actually takes place due to 
vanishing loop currents. Summarizing, we have in the limit $m_d\rightarrow\infty$,
\begin{equation}
\dot{W}_\infty= \left\{
\begin{array}{cc}
{2\over \beta\tau_m}(1-\alpha  )\ln (\alpha_{EQ}/\alpha), & 0<\alpha<1;\\
 0, & 1\le \alpha \le \alpha_{EQ}.
\end{array}\right.
\label{eq:dotW_inf}
\end{equation}
The singular behavior of $\dot{W}$ against $\alpha$ indicates a true nonequilibrium transition in the model
where the methylation range is infinite.

According to Eq. (\ref{eq:errorApprox}), the adaptation error $\epsilon$ can be made arbitrarily small in the
entire adaptive phase $\alpha<\alpha_c=1$ by increasing $m_d$. On the other hand, $\dot{W}$ can be
made arbitrarily small at the same time by choosing an $\alpha$ close to $\alpha_c$.
The energy dissipation is necessary to generate adaptive behavior, however, there does not appear to
be a minimal value for the dissipation rate to support an arbitrarily accurate adaptive system.  

For a system with a finite methylation range, transition between the two phases is more gradual than what
is described above. From Eqs. (\ref{eq:probability-active}) and (\ref{eq:probability-noadapt}), one may
identify a ``correlation length'' $\lambda\simeq 1/|\ln\alpha|\simeq |1-\alpha|^{-1}$.
For $m_d>\lambda$, Eq. (\ref{eq:dotW_inf})  can be directly applied. Corrections need to
be considered when $m_d<\lambda$, based on exact results derived in previous sections.

\section{Conclusions}
\label{sect:conclusion} 

In this paper, we report a detailed analytical study of the stochastic
network model proposed by Lan et al. shown previously to
describe well sensory adaptation in E. coli.   
To understand this system, we first derive moment equations which are closely related to the rate equations traditionally 
used to model this type of biological processes.  The moment equations implement an integral feedback control scheme 
at the heart of the adaptive behavior.  Adaptation in the model is achieved when the nonequilibrium parameter 
$\alpha<\alpha_c$, where $\alpha_c=1$ is less than its value $\alpha_{EQ}$ when detailed balance is observed. 
By mapping the original ladder network to a one-dimensional birth-death process under the assumption of 
timescale separation, we obtain analytic expressions for the NESS distribution with qualitatively different 
behavior for $\alpha>1$ (nonadaptive phase) and $\alpha<1$ (adaptive phase).   
With the help of the exact results on the NESS distribution, we compute the mean receptor activity from which
its dependence on the external signal strength is obtained. 
In the adaptive phase, the adaptation error, which measures the
deviation of the mean receptor activity from a suitable reference value, is found to decrease when the system
is driven further out of equilibrium by reducing $\alpha$, but approaches a saturated value for $\alpha<\alpha_m$.
Interestingly, at a given $\alpha$, the adaptation error decreases exponentially with the number of the available
methylation states before the extreme methylation levels are reached.

We also derive an exact formula for the energy dissipation rate by using a cycle-decomposition technique.  
The energy dissipation rate is found to be insensitive to the size of the methylation range and also to timescale separation.
Although our results confirm qualitatively the statement that adaptation within the molecular construct that
implements integral feedback control requires nonequilibrium driving, there does not appear to be a lower bound
on energy dissipation to achieve a given level of adaptation accuracy, in contrary to the 
Energy-Speed-Accuracy tradeoff relation proposed previously by Lan et al.

Although the methylation range of a single MCP receptor is four, we have investigated the behavior of the system with arbitrary methylation range, especially when the methylation range $m_0$ is large. The extension allows us to examine
various theoretical issues quantitatively. In the limit $m_0\rightarrow\infty$,
a true nonequilibrium phase transition at $\alpha=1$ can be identified. 

The current analysis only focuses on the static properties of the system. However, the transient response
at short times is also an important component of the molecular adaptive circuit. 
An understanding of the adaptive behavior in a general setting based on thermodynamic principles is still lacking.   
In this respect, lessons may be drawn from recent developments in
information thermodynamics~\cite{sartori2014thermodynamic,maruyama2009colloquium,parrondo2015thermodynamics}
by viewing the adaptive circuit as an information processing machine. 

\section*{Acknowledgement}
We thank David Lacoste, Kirone Mallick and Henri Orland  for helpful discussions. The work is supported in part by the Research Grants Council of the HKSAR under grant HKBU 12301514.

\appendix
\section{Approximate expression of the NESS distribution}

In this Appendix we derive an approximate analytic expression for the NESS distribution
given by Eq. (\ref{eq:coar-PDF}). Taking the logarithm of the equation, we obtain,
\begin{equation}
\ln {P(m+1)\over P(0)}= \sum_{i=0}^m \ln{b(i)\over d(i+1)} 
=\ln {d(0)\over d(m+1)}+\sum_{i=0}^m\phi\bigl(x(i)\bigr),
\label{eq:log_P}
\end{equation}
where 
\begin{equation*}
\phi(x)=\ln{1+\alpha e^x\over \alpha + e^x}
\end{equation*} 
and 
$x(m)\equiv -\beta\Delta E(s,m) +\ln(k_{CB}/k_{CR})=\beta e_0(m-m^*)$, with
$m^*=m_1+f(s)/e_0-(\beta e_0)^{-1}\ln(k_{CB}/k_{CR})$. It is straightforward to verify that $\phi(x)$ is 
an odd function of $x$.

The function $\phi(x)$ is well approximated by a piece-wise linear function 
\begin{equation}
\psi(x)=
 \left\{ 
\begin{array}{cc}
-\ln\alpha, &  x\leq -\xi;\\
{\alpha -1\over\alpha+1}x, & -\xi<x<\xi;\\
\ln\alpha,& x\geq\xi.
\end{array}\right.
\label{eq:phi-approx}
\end{equation}
which has the same slope at $x=0$ and same asymptotic values as $x\rightarrow\pm\infty$.
Continuity requires the choice $\xi={\alpha +1\over\alpha -1}\ln\alpha$. The two functions match each other
well except near $x=\pm\xi$. 

For $\alpha<1$, the peak of $P(m)$ is centered at $m^\ast$. It is thus convenient to 
use $P(m^*)$ as the reference. Approximating the sum in Eq. (\ref{eq:log_P})
by an integral over $\psi(x)$, we obtain,
\begin{equation}
\fl
\ln {P(m)\over P(m^*)}\simeq \left\{
\begin{array}{ll}
-(m-m^\ast)\ln\alpha + \ln{d(m^*)\over d(m)}
 + {1+\alpha\over 1-\alpha}{1\over 2\beta e_0}\ln^2\alpha, & m<m_-; \\
-{1-\alpha\over 1+\alpha}{\beta e_0\over 2}(m-m^\ast)^2 +  \ln{d(m^*)\over d(m)}, &m_-< m <m_+; \\
(m-m^\ast)\ln\alpha + \ln{d(m^*)\over d(m)}+{1+\alpha\over 1-\alpha}{1\over 2\beta e_0}\ln^2\alpha, &   m>m_+. 
\end{array}\right.
\label{eq:probability-active}
\end{equation}
Here $m_\pm=m^\ast\pm\xi/(\beta e_0)$. Since $d(m)$ has only a relatively weak dependence on $m$,
we see that $P(m)$ is essentially gaussian within the interval $(m_-,m_+)$, but turns to simple exponential
decay outside the interval.

For $\alpha>1$, $P(m)$ achieves its minimum value at $m^\ast$. In the neighborhood of the methylation
boundaries, we have
\begin{equation}
P(m)\simeq 
\left\{
\begin{array}{ll}
\alpha^{-m}P(0)d(0)/d(m), & m<m_-;\\
\alpha^{m-m_0}P(m_0)d(m_0)/d(m), & m>m_+.
\end{array}\right.
\label{eq:probability-noadapt}
\end{equation}
Noting that $\phi(x)$ is an odd function of $x$, we have approximately
$P(m_0)\simeq \alpha^{m_0-2m^\ast}P(0)d(0)/d(m_0)$.


 \end{document}